\documentclass[iop,apj,tighten]{emulateapj}

\slugcomment{Accepted for publication in the Astrophysical Journal} 

\usepackage{apjfonts}

\usepackage{graphicx}
\usepackage[breaklinks,colorlinks=true,linkcolor=blue,citecolor=blue,urlcolor=blue]{hyperref}
\usepackage{bm}
\usepackage{color}
\usepackage{enumitem}
\usepackage{amsmath}

\usepackage{CJK}

\newcommand{\mdeg}[1]{^\circ#1}
\newcommand{\ddeg}[1]{\mspace{-6.3mu}{^\circ{\mspace{-1.0mu}#1}}}

\usepackage{etoolbox}
\makeatletter
\patchcmd{\NAT@citex}
  {\@citea\NAT@hyper@{%
     \NAT@nmfmt{\NAT@nm}%
     \hyper@natlinkbreak{\NAT@aysep\NAT@spacechar}{\@citeb\@extra@b@citeb}%
     \NAT@date}}
  {\@citea\NAT@nmfmt{\NAT@nm}%
   \NAT@aysep\NAT@spacechar\NAT@hyper@{\NAT@date}}{}{}
\patchcmd{\NAT@citex}
  {\@citea\NAT@hyper@{%
     \NAT@nmfmt{\NAT@nm}%
     \hyper@natlinkbreak{\NAT@spacechar\NAT@@open\if*#1*\else#1\NAT@spacechar\fi}%
       {\@citeb\@extra@b@citeb}%
     \NAT@date}}
  {\@citea\NAT@nmfmt{\NAT@nm}%
   \NAT@spacechar\NAT@@open\if*#1*\else#1\NAT@spacechar\fi\NAT@hyper@{\NAT@date}}
  {}{}
\makeatother

\shorttitle{Polarity Reversal}
\shortauthors{Sun et al.}

\begin{document}

\begin{CJK}{UTF8}{}

\title{
On Polar Magnetic Field Reversal and Surface Flux Transport During Solar Cycle 24
}

\author{
\begin{CJK}{UTF8}{gbsn} 
Xudong Sun (孙旭东), J. Todd Hoeksema, Yang Liu (刘扬), and Junwei Zhao (赵俊伟)
\end{CJK}
}

\affil{W. W. Hansen Experimental Physics Laboratory, Stanford University, Stanford, CA 94305; 
\href{mailto:xudong@sun.stanford.edu}{xudong@sun.stanford.edu}}


\begin{abstract}
As each solar cycle progresses, remnant magnetic flux from active regions (ARs) migrates poleward to cancel the old-cycle polar field. We describe this polarity reversal process during Cycle 24 using four years (2010.33--2014.33) of line-of-sight magnetic field measurements from the Helioseismic and Magnetic Imager. The total flux associated with ARs reached maximum in the north in 2011, more than two years earlier than the south; the maximum is significantly weaker than Cycle 23. The process of polar field reversal is relatively slow, north-south asymmetric, and episodic. We estimate that the global axial dipole changed sign in October 2013; the northern and southern polar fields (mean above $60\mdeg$ latitude) reversed in November 2012 and March 2014, respectively, about 16 months apart. Notably, the poleward surges of flux in each hemisphere alternated in polarity, giving rise to multiple reversals in the north. We show that the surges of the trailing sunspot polarity tend to correspond to normal mean AR tilt, higher total AR flux, or slower mid-latitude near-surface meridional flow, while exceptions occur during low magnetic activity. In particular, the AR flux and the mid-latitude poleward flow speed exhibit a clear anti-correlation. We discuss how these features can be explained in a surface flux transport process that includes a field-dependent converging flow toward the ARs, a characteristic that may contribute to solar cycle variability.
\end{abstract}



\section{Introduction}
\label{sec:intro}

The large-scale solar magnetic field is dipole-like during the minimum phase of the activity cycle. Concentrations of flux in the polar region are predominantly unipolar with a mean field of several gauss \citep{svalgaard1978,tsuneta2008}. Away from the minimum, decaying active-region (AR) flux migrates poleward. The polar field and the dipole moment reverse sign around the maximum phase \citep{babcock1959}, which has been observed or inferred since the early twentieth century \citep{howard1972,makarov1983,webb1984,fox1998,durrant2003,hoeksema2010,mjaramillo2012}.

The polarity reversal in the photosphere may be described by a surface flux transport (SFT) process that considers the radial field ($B_r$) evolution \citep[e.g.,][]{wangym1989sci}. Therein, magnetic flux is introduced through bipolar AR emergence and dispersed by differential rotation, meridional flow, and supergranular diffusion. Because the preceding-polarity component of an AR is generally at a lower latitude compared to the trailing \citep[Joy's law on AR tilt,][]{hale1919}, its flux preferentially diffuses across the equator and cancels with its opposite hemisphere counterpart. This leaves a net trailing polarity flux being transported poleward to cancel the old polar field. It it notable that the net flux in the polar region during minimum phase is comparable to the unsigned flux in a single major AR (several $10^{22}$~Mx).

\begin{figure}[t!]
\centerline{\includegraphics{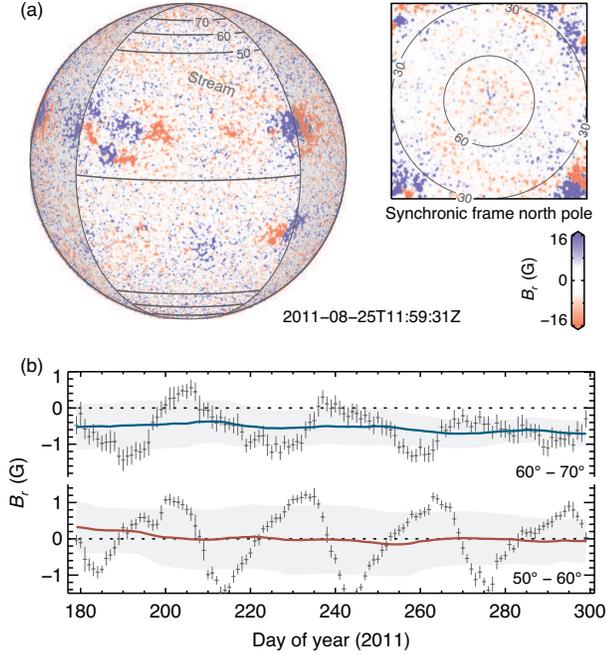}}
\caption{Sample HMI observation. (a) Inferred radial field ($B_r$) magnetogram. Data within $\pm45\mdeg$ longitude (unshaded) are used for mean field estimation at each latitude. A negative flux stream (red) extends toward the northern polar region. Inset: stereographic view of the north pole from a co-temporal synchronic frame. The sector in the bottom third is updated with new data.  A four-year sequence is available as an online animation. (b) Mean $B_r$ in the indicated latitudinal bands. Data points represent daily means. Here and after, curves show area-weighted running average for the full Sun; error bars are weighted s.d. Shaded bands represent s.d. of all single-magnetogram means within the running-average window. (An animation of the figure is available at \url{http://goo.gl/VtyDwG}.)}
\label{f:sync}
\end{figure}

\begin{figure}[t!]
\centerline{\includegraphics{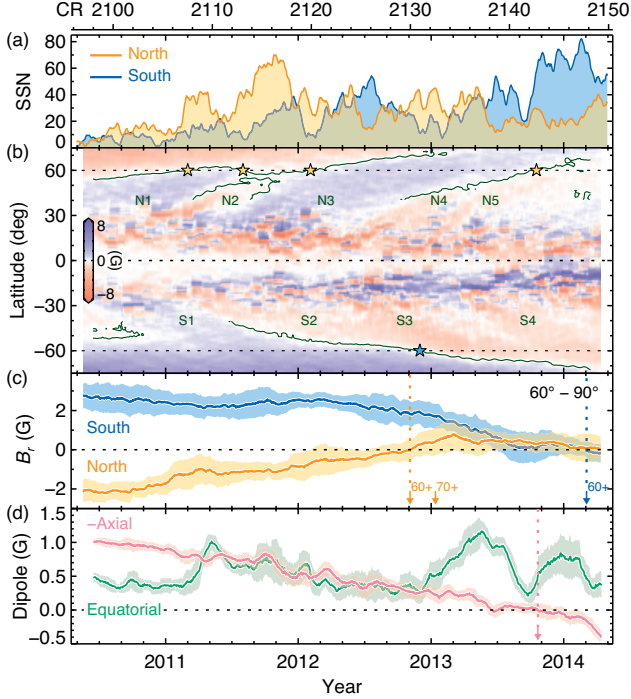}}
\caption{Magnetic field evolution. (a) SIDC hemispheric sunspot number (SSN). (b) Time-latitude diagram of zonally averaged $B_r$. Contours indicate polarity inversion; their intersections with horizontal dotted lines indicate reversals at $\pm60\mdeg$ (star symbols). Here and after, N1--N5 and S1--S4 mark individual flux surges in the northern and southern hemispheres. (c) Mean $B_r$ above $60\mdeg$ as polar fields. (d) Global axial and equatorial dipole. The axial component is multiplied by $-1$ for better comparison. Averaging window is $30\mathrm{~days}$ for (a) (d). Vertical dotted lines and arrows mark the reversal times of polar fields above the denoted latitudes (c), and axial dipole (d).}
\label{f:rev}
\end{figure}

The SFT process is evidenced by the unipolar flux streams that extend to high latitudes from the decaying ARs (Figure~\ref{f:sync}). Averaged zonally, they appear as poleward surges of flux in the time-latitude diagram (Figure~\ref{f:rev}(b)). The surges can be north-south symmetric or anti-symmetric at times, but are more often asymmetric and episodic \citep{howard1981,hoeksema1991,svalgaard2013}.

Polarity reversal is an integral part of the Babcock-Leighton (B-L) solar cycle mechanism \citep{babcock1961,leighton1969}. Emerging tilted ARs provide flux for the new-cycle poloidal field, whereas large-scale flows and diffusion redistribute it and control the efficiency with which it coalesces to global scale \citep{wangym1991}. These elements exhibit intra- and inter-cycle variations \citep{basu2010,hathaway2010,despuig2010,ulrich2010}. SFT modeling suggests that the variations in sources and in flux transport parameters can have long-term effects \citep{wangym2002,baumann2004,cameron2010,jiang2013}, providing clues to solar cycle variability. In particular, a field-dependent inflow toward the activity belts \citep{gizon2001,haber2004,zhaojw2004} is thought to impose nonlinear feedback to the process \citep{jiang2010,cameron2012}.

The ongoing Cycle 24 is preceded by an unusually long and quiescent minimum. Its evolution provides an interesting opportunity to scrutinize various solar cycle models \citep{pesnell2012}. Here, we study its polarity reversal and implications on the SFT process using data from the Helioseismic and Magnetic Imager \citep[HMI;][]{schou2012} on board the \textit{Solar Dynamic Observatory} (\textit{SDO}). The high-cadence, high-duty-cycle, moderate-resolution observations allow continuous, high-accuracy measurement of the polar fields through deep averaging, and enable a systematic probe of the near-surface flows through local helioseismology.

Several studies have examined aspects of the current reversal \citep[e.g.,][]{shiota2012,upton2014,benevolenskaya20014,karna2014}. We seek to characterize the mean field evolution at various latitudes, and demonstrate the influence of several factors (mean AR tilt, total AR flux, and meridional flow) on the flux surges. We discuss the results in the SFT/B-L context, inclusive of the aforementioned AR inflow.


\section{Data Reduction}
\label{sec:data}

We make use of the HMI 720-s cadence line-of-sight (LOS) magnetograms ($B_l$). The dataset includes $\sim$$1.61\times10^5$ good-quality images between 2010 May 1 and 2014 April 30 (Carrington rotation, CR 2096--2150). We use data within $\pm45\mdeg$ longitude of central meridian and 0.995 disk radii, and assume that the field vectors are radially oriented \citep{svalgaard1978,wangym1992}, $B_r=B_l \mu^{-1}$, where $\mu$ is the cosine of angle between the LOS and the local normal (Figure~\ref{f:sync}(a)). The radial-field assumption has been widely used in data analysis and modeling with success. Future study with HMI full-disk vector magnetograms \citep{hoeksema2014} can provide additional constraints on $B_r$. We discuss the prospect at the end of Section~\ref{sec:discussion}.

For each magnetogram, we calculate the weighted mean $B_r$ for each latitudinal bin, $\sum w B_r  / \sum w$, where $w\propto \mu^{-1}$ denotes the de-projected magnetogram pixel area. To estimate the full-Sun mean, we further perform an area-weighted running average. For each latitude, the temporal width corresponding to the entire Sun is determined by a differential rotation profile, $26.8\mathrm{~days}$ at the equator and $38.5\mathrm{~days}$ at the poles. Thus the number of magnetograms averaged varies from $3.2\times10^3$ to $4.6\times10^3$, from equator to pole. We report the result with one-day resolution. Our approach utilizes a wider longitudinal window compared to the traditional synoptic maps. The moderate-resolution data and deep averaging reduce the large temporal variations when the pole is tilted away. The result is not too sensitive to the running-average window size.

The statistical uncertainty for a 720-s LOS magnetogram measurement in one pixel is about $6.3~\mathrm{G}$ \citep{liuy2012}; it is expected to vary as $\mu ^{-1}$ for the inferred $B_r$. We evaluate the weighted standard deviation (s.d.) of $B_r$ in the northern polar region (above $60\mdeg$) for each magnetogram for 2011, and find the 30-day median ranging between $23.5\mathrm{~G}$ and $41.4\mathrm{~G}$. The variation in the running average is dominated by magnetic features (ARs and flux streams) moving in and out of the longitudinal window (Figure~\ref{f:sync}(b)).

We consider the zero crossings of mean $B_r$ to be the polarity reversal times. They are generally several months delayed compared to $B_l$, because the $\mu^{-1}$ factor in $B_r$ places greater weights on high latitudes, favoring the old-cycle polarity. Conversely, if the reversal occurs when the pole tilts away, the reversal is expected to be observed early. This is because the highest latitudes are not observed, so the mean is biased toward the new-cycle polarity from the lower latitudes.

To estimate the global dipole, we use the HMI synchronic frames, which update the synoptic maps with new observations each day (Figure~\ref{f:sync}(a)). Specifically, a third of the map is replaced by recent observations at the same Carrington longitude that are one-rotation newer. The well-observed polar data obtained in each spring or fall are interpolated to estimate the $B_r$ above $75\mdeg$ latitude at any given time and the smoothed, interpolated values are used to fill in the regions with data missing due to the unfavorable viewing angle \citep{sun2011}. We use a direct integration method for spherical harmonic decomposition. The $(l,m)=(1,0)$ and $(l,m)=(1,\pm1)$ constituents represent the axial and equatorial dipoles, respectively. The total dipole is the strength of their vector sum.

We evaluate the flux-weighted AR centroid latitude ($\lambda_{+}$ and $\lambda_{-}$) separately for each polarity in each hemisphere. We use their separation, $\Delta \lambda \propto (\lambda_{+} - \lambda_{-})$, as a proxy for the mean AR tilt angle. The value is computed each day for the entire Sun, using the running-average method described above for $B_r$. When the trailing polarity is higher, $\Delta \lambda$ is positive (normal tilt); otherwise it is negative (inverted tilt). This measurement avoids the uncertainty in sunspot identification, and naturally accounts for the polarity information. When many ARs are present, the majority follow Joy's law, so $\Delta \lambda$ is generally greater than zero. During low magnetic activity, $\Delta \lambda$ can be determined by a single AR complex. We include only pixels between $0\mdeg$--$40\mdeg$ latitude with $|B_r|>120\mathrm{~G}$. Such an empirical threshold appears to provide a good separation between the ARs and the flux streams. We also compute the total unsigned AR flux ($\Phi$) from these pixels, which mainly include sunspots and plages.

The product of the AR flux and the polarity separation, $P=\Phi \Delta \lambda$, has been used as a proxy for the AR poloidal field \citep{petrie2012}. It contributes to the new-cycle global poloidal field when $\Delta \lambda>0$, and is of fundamental importance to the B-L mechanism. Other methods for estimating $P$ exist too \citep[e.g.,][]{ulrich2002} but are not discussed here.

Cycle 24 started before the \textit{SDO} launch. For context, we also include data from January to May 2010 (CR 2092--2096) from the Michelson Doppler Imager \citep[MDI;][]{scherrer1995} in some of our analyses. We use only the $B_r$ synoptic maps. The values are empirically scaled by 0.71 to match HMI \citep{liuy2012}.

The HMI time-distance helioseismology pipeline utilizes Doppler measurements to infer the subsurface flow field. The flow maps extend to $\pm60\mdeg$ latitude and are available every eight hours. We employ the near-surface meridional flow synoptic maps in the depth of $0$--$1\mathrm{~Mm}$ ($u_y$), and subtract a four-year mean profile ($\bar{u}_y$) to reveal the residuals ($\delta u_y$), or flow variations. Positive (negative) $\delta u_y$ in the north (south) indicates faster poleward flow. The helioseismic results are in general agreement with those derived from surface feature tracking methods \citep{svanda2007}. We refer to \cite{zhaojw2014} for detailed processing procedures, flow characterization, and uncertainty estimation.


\begin{figure}[t!]
\centerline{\includegraphics{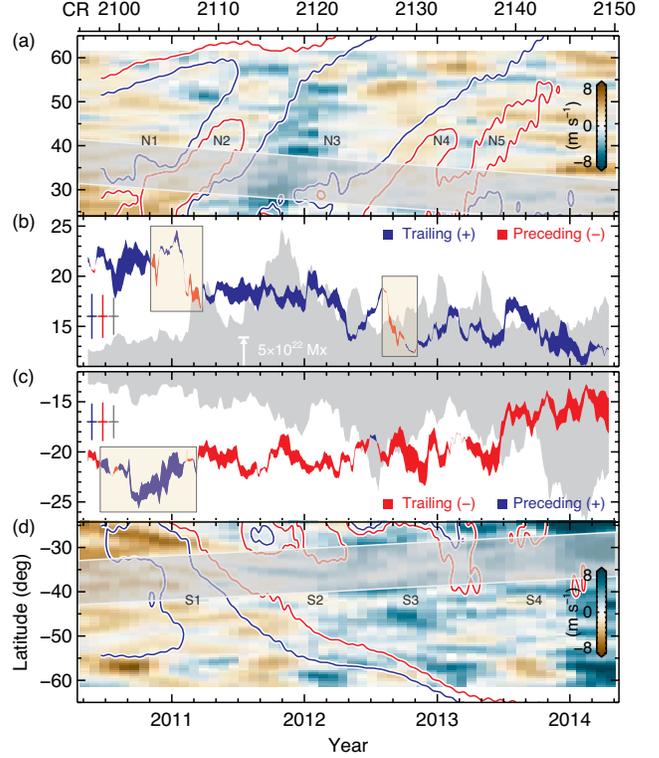}}
\caption{Evolution of flux surges and relevant factors. Panels (a) (d): Time-latitude diagram of zonally-averaged residual meridional flow $\delta u_y$ at mid latitude. Brown indicates faster poleward flow, green slower. Over-plotted are $B_r$ contours at $\pm0.3\mathrm{~G}$ (blue/red) that outline the flux surges. Shaded bands indicate $10\mdeg$--$20\mdeg$ poleward of the mean activity belt latitude $\lambda_c$; mean $B_r$ and $\delta u_y$ in this band are used to construct Figure~\ref{f:corr}. Panels (b) (c): Temporal profile of AR centroid $(\lambda_{+}+\lambda_{-})/2$. The curve width shows the polarity separation $\Delta \lambda$, a proxy of mean AR tilt. For north, blue (red) indicates positive (negative) $\Delta \lambda$, or that the trailing (preceding) polarity is higher. For south it is the opposite. Boxes highlight periods with smaller or inverted tilt. Gray shades show AR flux $\Phi$. Crosses in (b) (c) show median s.d. of $\lambda_{+}$ (blue), $\lambda_{-}$ (red), and $\Phi$ (grey); the relatively large value is owing to the fact that we only measure a quarter of the Sun each time. The standard errors of the mean are small as we include $3.3\times10^3$ measurements in each running average.}
\label{f:ftrans}
\end{figure}


\section{Results}
\label{sec:result}


\subsection{A North-South Asymmetric, Episodic Reversal}
\label{subsec:asym}

Cycle 24 started with negative polar field and has negative preceding sunspots in the north. The sunspot number (SSN) peaked late in 2011 in the north, more than two years earlier than the south (Figure~\ref{f:rev}(a)). The maximum SSN for north and south (30-day mean) is 72 and 80, respectively. The maximum total SSN is 105, only about $60\%$ of Cycle 23.

In the north, two large positive surges N1 and N3 contributed much to the polar field reversal; smaller negative surges N2, N4, and N5 interrupted the process (Figure~\ref{f:rev}(b)). The pattern became fragmented after N5. The polarity alternation resulted in multiple reversals around $60\mdeg$. In the south, a positive surge S1 first replenished the old-cycle polar field. Subsequent negative surges S2--S4 initiated and strengthened the reversal. Incidences of positive flux separated S2--S4. These are consistent with the SOLIS observations \citep{petrie2014}.

The north-south asymmetric surges led to asymmetric polar field reversals (Figure~\ref{f:rev}(c)). The northern polar field (mean $B_r$ above $60\mdeg$) gradually decreased in strength from about $-2\mathrm{~G}$ and switched sign in November 2012. The growth of the new polar field, however, was stalled due to old-polarity surges N4 and N5. An old-cycle polarity comeback and more reversals are expected. The southern polar field remained between $2$ and $3\mathrm{~G}$ until mid-2012 owing to S1. It weakened afterwards, paused near zero during late-2013, and finally changed sign in March 2014.

The reversal times differ by approximately 16 months; both polar fields were positive and close to zero for over a year. For $70\mdeg$ and above, the north reversed in January 2013; the south had not by May 2014. The early reversal in the north was accompanied by negative (preceding-polarity) flux diffusing across the equator prior to mid-2013 (Figure~\ref{f:rev}(b)). The low-latitude region rapidly turned positive after November 2013 owing to the rising activity in the south.

During the polar field reversal, the total dipole did not vanish and remained above $0.2\mathrm{~G}$ (Figure~\ref{f:rev}(d)). From early-2011 to late-2012, the axial and equatorial dipole strengths were similar; both gradually decreased at about $0.3\mathrm{~G~yr}^{-1}$. The equatorial component subsequently peaked in phase with the rising SSN in the south. The axial dipole stayed below $0.1\mathrm{~G}$ for the latter part of 2013 and displayed multiple zero crossings. It changed sign definitively in October 2013, then recovered quickly to exceed the equatorial counterpart in March 2014.


\begin{figure}[t!]
\centerline{\includegraphics{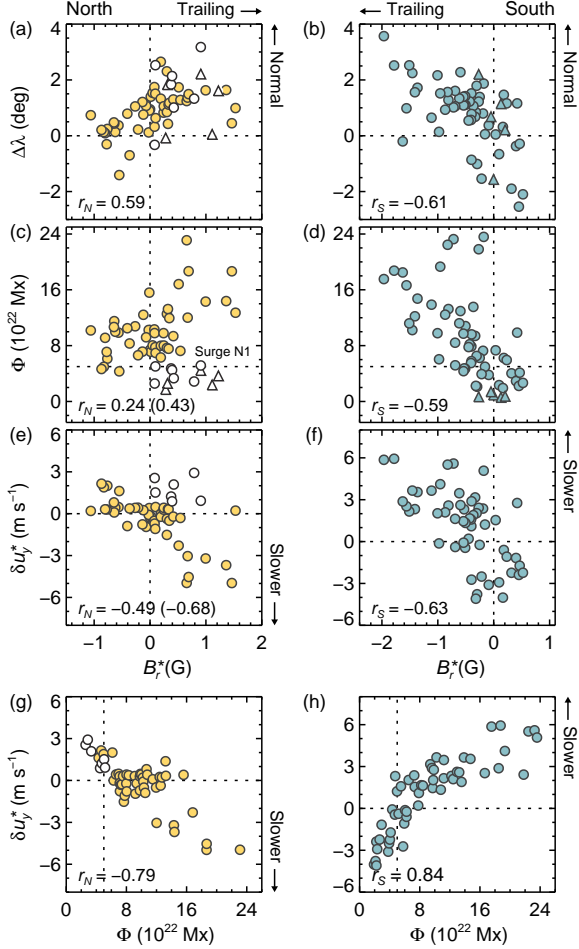}}
\caption{Correlation between various constituents of the SFT process: mean AR tilt $\Delta \lambda$, AR flux $\Phi$, surge field $B_r^{\star}$, and residual meridional flow velocity $\delta u_y^{\star}$. Left is for north; right is for south. Circles show CR-means from HMI; triangles are from MDI. Open symbols indicate surge N1, which deviates from the overall trend in (c) and (e).  $B_r^{\star}$ and $\delta u_y^{\star}$ are from the latitudinal bands in Figure~\ref{f:ftrans}. Pearson's correlation coefficient $r$ is noted (HMI only); values in parentheses have the outlier N1 excluded. Probability of chance occurrence as significance: $p=0.068$ for (c); $p<10^{-4}$ for others. All standard errors of the mean are small.}
\label{f:corr}
\end{figure}


\subsection{What Influence the Flux Surges?}
\label{subsec:surge}

SFT modeling show that the variations of mean AR tilt, total AR flux, and meridional flow can have long term effects  \citep{wangym2002,baumann2004,cameron2010,jiang2013}. Here, we search for the observational signatures of how these factors influence the flux surges. We defer interpretation to Section~\ref{sec:discussion}.

To spatially separate ARs and flux surges, we fit a linear function of time $\lambda_c(t)=k(t-t_0)+\lambda_{c}(t_0)$ to the AR centroid latitude $(\lambda_{+}+\lambda_{-})/2$ in each hemisphere and define it as the activity belt mean latitude. Here, $t_0=2010.33$ denotes our reference time, 2010 May 1. Over four years, $\lambda_c$ in the north moved equatorward at a rate of $k_N=-2.\ddeg30\pm0.\ddeg04~\mathrm{yr}^{-1}$;  in the south the rate is $k_S=1.\ddeg58\pm0.\ddeg03~\mathrm{yr}^{-1}$. As of 2014 April 30, $\lambda_c$ was $12.\ddeg2$ and $-16.\ddeg6$ for north and south, respectively. As most sunspots are located within $10\mdeg$ of $\lambda_c$, we use the values in the latitudinal band $10\mdeg$--$20\mdeg$ poleward of $\lambda_c$ to represent the flux surge condition (Figures~\ref{f:ftrans}(a)(d)).

We denote the magnetic field and the residual meridional flow velocity in this band as $B_r^{\star}$ and $\delta u_y^{\star}$, respectively. The influence of ARs on the surges may be quantified by determining the correlation between the mean tilt proxy $\Delta \lambda$, the total flux $\Phi$, and $B_r^{\star}$, $\delta u_y^{\star}$. The putative influence should be strong immediately poleward of the activity belt and become weaker with a time lag at higher latitudes owing to dispersive processes and finite poleward transport time.

\textit{Mean AR tilt.}---The mean AR tilt proxy $\Delta \lambda$ is usually positive, indicating that the majority of ARs possess normal tilt in accordance with Joy's law (Figures~\ref{f:ftrans}(b)(c)). The magnitude of $\Delta \lambda$ is a few degrees; the root-mean-square (r.m.s.) is $1.\ddeg35$. Three boxes in the figure highlight three extended periods of small or negative $\Delta \lambda$ (inverted tilt) in the profile; they correspond to the origins of the preceding-polarity surges (N2, N4/N5, and S1). Intervals of positive $\Delta \lambda$ (normal tilt), on the other hand, generally correspond to the trailing-polarity surges.

We quantify this trend by correlating the CR-mean $B_r^{\star}$ and $\Delta \lambda$, and find the following correlation coefficients: $r_N=0.59$ in the north, and $r_S=-0.61$ in the south\footnotemark[1] (Figures~\ref{f:corr}(a)(b)). This indicates that the higher-latitude AR component tends to be transported poleward regardless of its polarity, and that greater initial AR polarity separation leads to stronger field in the surges.

\footnotetext[1]{Here and after, $r_N$ and $r_S$ have the same physical meaning, but have opposite sign because of the opposite-signed trailing-polarity $B_r$, as well as our particular definitions of $\Delta \lambda$ and $\delta u_y$.}

\textit{Total AR flux.}---The total AR flux $\Phi$ and SSN are well correlated, with $r=0.96$. The hemispheric AR flux $\Phi$ varied significantly with multiple peaks and troughs (Figures~\ref{f:ftrans}(b)(c)). The maximum is about $2.4\times10^{23}\mathrm{~Mx}$ and $2.6\times10^{23}\mathrm{~Mx}$ for north and south, respectively. Similar to SSN, total flux in the north peaked more than two years earlier than the south. 

\begin{figure}[t!]
\centerline{\includegraphics{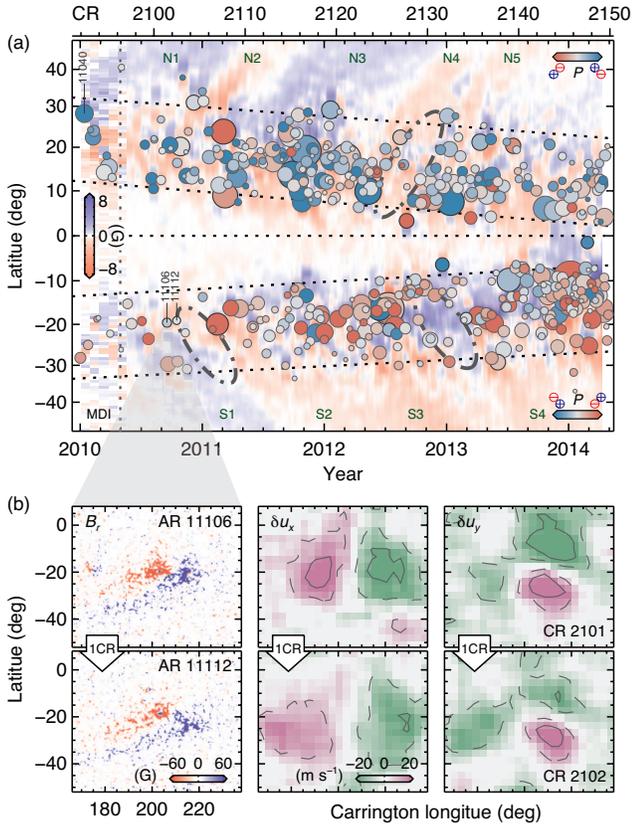}}
\caption{Selective AR properties of Cycle 24. (a) Sunspot butterfly diagram constructed from the DPD sunspot catalogue \citep{gyori2011} over-plotted on $B_r$. Magnetic data before May 2010 are from MDI. Each circle represents a NOAA AR with umbrae larger than $10^{-5}$ solar hemisphere surface area. The symbol size is proportional to the maximum sunspot area within $\pm30\mdeg$ of central meridian. The color indicates a poloidal field proxy for individual ARs, $P \propto {\Phi d \sin \alpha} \propto d^{2.3}\sin \alpha$, where $d$ is the polarity separation and $\alpha$ the tilt, with $\Phi \propto d^{1.3}$ \citep{wangym1989soph}. For north, ARs with normal tilt are blue; inverted tilt, red. For south it is the opposite. A blue symbol is expected to contribute positive flux to the poleward surge. ARs with inverted tilt do exist when the surge field is of the trailing sunspot polarity. Ellipses denote regions devoid of major ARs. Inclined dotted lines show the band $\lambda_c\pm10\mdeg$. (b) Inflows around one recurrent region AR 11106 (top) and 11112 (bottom). From left to right, panels show parts of synoptic maps of $B_r$, $\delta u_x$ (residual zonal flow velocity), and $\delta u_y$. The converging flow pattern is extensive and field-dependent. Contours are at $\pm$10 and $\pm20\mathrm{~m~s}^{-1}$.}
\label{f:ars}
\end{figure}

The sunspot butterfly diagram provides an intuitive view of the latitudinal and temporal distribution of ARs (Figure~\ref{f:ars}(a)). There are three relatively quiescent periods as indicated by the dashed-dotted ellipses, which overlap with the preceding-polarity surges (N4/N5, S1, and the gap between S3 and S4). Higher AR flux, on the other hand, overlapped with the trailing-polarity surges (N3, S3, and the upcoming one after S4). We find a weaker $B_r^{\star}$--$\Phi$ correlation: $r_N=0.24$, $r_S=-0.59$ (Figures~\ref{f:corr}(c)(d)). The correlation is weak in the north mainly because of the counterexample surge N1, which has low AR flux and trailing-sunspot polarity. As shown in Section~\ref{sec:discussion}, this provides interesting clues that help explain the observed trends. The correlation improves to $r_N=0.43$ if we exclude N1.

\textit{Meridional flow.}---Between $30\mdeg$ and $60\mdeg$ latitude, the residual meridional flow velocity $\delta u_y$ has a r.m.s. value (for box-car means of $1\mathrm{~CR}\times2\mdeg$) of $1.7\mathrm{~m~s}^{-1}$ and $2.5\mathrm{~m~s}^{-1}$ for north and south. The relative variation with respect to the mean profile, $\delta u_y / \bar{u}_y$, is 0.21 and 0.37 for north and south, respectively. This demonstrates that the mid-latitude meridional flow varies significantly over the course of a solar cycle.  

The time-latitude diagram of $\delta u_y$ indicates that the mid-latitude poleward flow was faster than average before 2011 (Figures~\ref{f:ftrans}(a)(d)). Remarkably, the flow only slowed down at the surge boundaries where $B_r$ switches polarity (N2/N3 and S1/S2). Coherent slower flow appeared within the trailing-polarity surges (N3 and S3). It seems that the meridional flow speed is curiously dependent on the sign of the magnetic field \citep{zhaojw2014}.

Correlating $B_r^{\star}$ and $\delta u_y^{\star}$ yields $r_N=-0.49$, and $r_S=-0.63$ (Figures~\ref{f:corr}(e)(f)). The result is similar if we use a different latitudinal band, or study the overall velocity $u_y$ instead of $\delta u_y$ \citep{zhaojw2014}. This indicates that the trailing-polarity (preceding-polarity) flux tends be transported when the poleward flow is slower (faster).  Surge N1 is again an exception, with faster poleward flow and trailing-sunspot polarity. The correlation increases to $r_N=-0.68$ if N1 is excluded.

\textit{Meridional flow vs. AR flux.}---We find a tighter $\Phi$--$\delta u_y^{\star}$ correlation: $r_N=-0.79$, $r_S=0.84$ (Figures~\ref{f:corr}(g)(h)). The overall anti-correlation between $\Phi$ and the residual poleward flow speed is $r=-0.81$. That is, the poleward flow is slower (faster) when the AR flux is higher (lower). The anti-correlation holds throughout mid latitudes. Surge N1 fits the trend too.


\section{Discussion and Summary}
\label{sec:discussion}

Cycle 24 has been weak, consistent with predictions based on the weak polar field late in Cycle 23 \citep[e.g.,][]{svalgaard2005,choudhuri2007}. Data from the \textit{Wilcox Solar Observatory} show that the polar field decline was slower than the last three cycles \citep[see][]{hoeksema2010}.

The asymmetric polar field reversal is not unusual though: the south led by 18 months during Cycle 19 \citep{babcock1959}, whereas the north has led by a year or so for all subsequent cycles \citep{svalgaard2013}. Magnetic activity too is generally north-south asymmetric: historical SSN records suggest that the hemispheric peaks can be offset by up to two years, a phase lag of 0.2 \citep{temmer2006,norton2010}. These two aspects are closely linked in light of the SFT process. The north-south asymmetries have far-reaching consequences throughout the heliosphere \citep[e.g.,][]{mcintosh2013}.

Section~\ref{subsec:surge} demonstrates that flux surges of the trailing (preceding) sunspot polarity are often accompanied by normal (inverted) mean AR tilt, higher (lower) AR flux, and slower (faster) meridional flow, with exceptions during low magnetic activity (e.g., surge N1). In the B-L framework, $\Delta\lambda$ and $\Phi$ determine the source poloidal field ($P \propto \Phi \Delta \lambda$), while $\delta u_y$ and diffusion (not studied here) modulate its redistribution. We discuss their roles in shaping the observed trends.

\textit{Source.}---The surge field $B_r^{\star}$ is correlated with AR poloidal field $P$ ($r_N=0.66$, $r_S=-0.62$); a tight correlation exists over the last 40 years \citep{petrie2012}. This is illustrated in Figure~\ref{f:ars}(a), where ARs in general appear to contribute to the surge the same sign of flux as their poloidal field. We note that the flow elements that form the surge are subject to dispersion processes, so the polarity of the surge does not necessarily agree in sign with $P$ of individual ARs in the figure. Apparent sign conflicts for some regions may also result from the nature of the DPD catalogue and our choice of representation, e.g., anti-Hale ARs with wrong-signed $P$; the large uncertainty in both $d$ and $\alpha$ for fast-evolving ARs; the AR flux obtained from empirical scaling rather than direct measurement; or contributions from multiple smaller ARs that are not included in the plot.

The $B_r^{\star}$--$\Delta \lambda$ relation (Figures~\ref{f:corr}(a)(b)), i.e., the preferential poleward transport of the higher-latitude component, is consistant with the preferential transequatorial cancellation of the lower-latitude counterparts irrespective of polarity.

Regarding the $B_r^{\star}$--$\Phi$ relation (Figures~\ref{f:corr}(c)(d)), we note that the mean tilt of a larger number of ARs should by statistics more robustly reflect Joy's law. The well-determined tilts lead to a positive and stable $\Delta \lambda$, so higher $\Phi$ consistently yields greater positive $P$, and consequently stronger trailing-polarity $B_r^{\star}$. ARs with inverted tilt do exist during these periods \citep[Figure~\ref{f:ars}(a);][]{mcclintock2013}; however, their contribution is relatively small. Conversely, $\Delta \lambda$ will have larger scatter when sunspots are few (lower $\Phi$). The source poloidal field $P$ becomes more random, so a similar $B_r^{\star}$--$\Phi$ trend does not necessarily hold.

We note that individual ARs with large tilt of \textit{either} sign can greatly influence global conditions, especially during intervals of low magnetic activity (e.g., AR 11040 for surge N1 and 11106/11112 for S1 as marked in Figure~\ref{f:ars}). The scatter of $\Delta \lambda$ reflects the randomness during the toroidal-poloidal field conversion in the B-L process, which may be important in shaping the cycle. Indeed, surge N1 canceled about half of the old polar flux within just one year in the early ascending phase of Cycle 24 (Figure~\ref{f:rev}). If these early ARs had smaller tilts, the reversal in the northern hemisphere might have been much delayed.

\textit{Modulation.}---Numerical SFT experiments show that a faster \textit{background} meridional flow profile (when above a certain threshold) leads to a weaker polar field \citep{baumann2004}. Intuitively, a faster flow rapidly sweeps both polarities poleward: transequatorial flux cancellation becomes ineffective, and further polarity separation is suppressed \citep{wangym2002}. Slower flow has the opposite effect, and more net flux will be brought poleward.

Unfortunately, this modulation mechanism does not differentiate between two polarities, so it alone does not explain why a trailing-sunspot polarity surge and slower poleward flow tend to appear together (Figures~\ref{f:corr}(e)(f)).

The fact that surge N1 appears as an outlier in both $B_r^{\star}$--$\Phi$ and here motivates us to use the more universal $\Phi$--$\delta u_y^{\star}$ relation as a link. Therein, slower poleward flow only appears during higher $\Phi$ (Figures~\ref{f:corr}(g)(h)), which leads to a positive source $P$ and trailing-polarity $B_r^{\star}$ as argued above. In other words, higher $\Phi$ ensures the concurrence of slower poleward flow and trailing-polarity field. Conversely, faster poleward flow is accompanied by lower $\Phi$, so $P$ and $B_r^{\star}$ are expected to vary (e.g., surge N1). This could explain the polarity dependence of meridional flow speed. It perhaps suggests that the $B_r^{\star}$--$\delta u_y^{\star}$ relation is largely determined by the AR properties, while the meridional flow modulation is of secondary importance.

\textit{AR inflow.}---The question then becomes, what causes the observed $\Phi$--$\delta u_y^{\star}$ trend, i.e., the anti-correlation between the AR flux and the mid-latitude poleward meridional flow speed? In addition to the meridional flow speed decrease on the poleward side of the activity belt, we also observe a speed increase on the equator-side of the activity belt during high magnetic activity. Such a converging pattern, visible to a depth of about $13\mathrm{~Mm}$ \citep{zhaojw2014}, likely reflects a near-surface convergent flow toward the ARs \citep[e.g.,][]{gizon2001,haber2004,zhaojw2004}, which is stronger and more extended for larger AR complexes (Figure~\ref{f:ars}(b)). The phenomenon is possibly a consequence of increased radiative loss due to strong magnetic field \citep{spruit2003,gizon2008}.

A feedback mechanism naturally arises: higher $\Phi$ leads to greater positive $P$, but also induces stronger converging flow, which effectively reduces the polarity separation, or $P$ \citep{jiang2010}. The dependence of the meridional flow on the AR field provides nonlinearity to the process. The inclusion of such a \textit{local} perturbation improves the SFT modeling of the cycle amplitudes \citep{cameron2012}. We note that the argument does not preclude a \textit{background} speed change \citep{gonzalez2008}. Prescribing faster background meridional flows during stronger cycles actually improves the SFT modeling of the open flux \citep{wangym2002}. More work is required to disentangle the global and local effects.

We finally note that the radial-field assumption, $B_r=B_l \mu^{-1}$, was originally inferred from the evolution of LOS field at relatively low resolution \citep{svalgaard1978,wangym1992}. With the advent of new instrumentations such as HMI, it is ultimately desirable to utilize higher resolution vector observations to better constrain $B_r$. To this end, we perform a preliminary comparison between the 720-s HMI LOS and vector data. In unipolar flux patches in the polar region, the $B_l$ component of the vector field exhibits excellent agreement with the LOS data \citep[see also,][]{hoeksema2014}. The nominal $B_r$ from vector data, however, appears stronger than that obtained using the LOS data and the radial-field assumption. Preliminary analysis suggests that field vectors in these flux patches deviate from the radial direction toward the poles.

Previous studies have found systematic deviation from the radial direction either towards \citep{ulrich2013} or away from \citep{gosain2013} the poles. The deviation is expected to be small, otherwise a large annual modulation would appear in the inferred $B_r$ profile (cf. Figure~\ref{f:rev}(c)). Any resolution using vector measurements must take into account the two following issues. Firstly, the vector field can be meaningfully retrieved \textit{only} in these patches where the polarization signal is high. In the majority quiet regions, spectropolarimetric inversion tends to overestimate the transverse field owing to the low signal-to-noise ratio \citep{borrero2012}. An estimate of the mean polar field is difficult because it is unclear how much these quiet regions contribute. Secondly, the magnetic filling factor $\alpha$ is fixed at unity in the current HMI inversion scheme. Given HMI's limited spectral resolution, only the product of $\alpha$ and $B$ is well constrained \citep{centeno2014}. The unity filling factor is a good approximation for sunspots, but is not ideal for smaller polar magnetic flux patches \citep{tsuneta2008}, which will affect the inferred inclination angle \citep{lites1990}. We defer a comprehensive analysis to a future paper.

To conclude, we find Cycle 24 to be weak and asymmetric, in agreement with the notions that the polar field during the activity minimum is a good predictor of the subsequent maximum, and that north-south asymmetry is a common feature of the solar cycle.

Furthermore, we characterize the interdependence of various constituents of the SFT process. All observed correlations and counterexamples are consistent with dispersion of the AR poloidal field by evoking Joy's law and a field-dependent AR inflow. The observed surge field originates from the AR poloidal field, and is further modulated by the meridional flow with feedback from the AR inflow.

With the availability of HMI meridional flow observations, it is now possible to correlate the flow variation against the AR magnetic field in a spatially resolved manner. Data-driven SFT modeling with such additional input may provide new insights to the cycle modulation mechanism.


\acknowledgments
We thank the anonymous referee and R. Ulrich for constructive comments. We are grateful to X. Zhao for the spherical harmonic decomposition code, and A. A. Norton and B. McClintock for bringing the DPD catalogue to our attention. This work is supported by NASA contract NAS5-02139 (HMI). The HMI data are courtesy of NASA and the \textit{SDO}/HMI team. The SSN records are courtesy of WDC-SILSO, Royal Observatory of Belgium, Brussels.

{\it Facilities:} \facility{\textit{SDO}}.\\


\end{CJK}

\bibliographystyle{yahapj}
\bibliography{reversal}





\end{document}